\documentclass[12pt]{iopart}
\usepackage{epsfig}

\def\d{{\rm d}}

\def\lsim{\mathrel{\raise.3ex\hbox{$<$\kern-.75em\lower1ex\hbox{$\sim$}}}}
\def\gsim{\mathrel{\raise.3ex\hbox{$>$\kern-.75em\lower1ex\hbox{$\sim$}}}}

\def\cmm2{{\,\rm cm^{-2}}}
\def\cm2{{\,{\rm cm}^2}}
\def\cmm3{{\,{\rm cm}^{-3}}}
\def\gcmm3{{\,{\rm g\,cm^{-3}}}}

\def\fun#1#2{\lower3.6pt\vbox{\baselineskip0pt\lineskip.9pt
  \ialign{$\mathsurround=0pt#1\hfil##\hfil$\crcr#2\crcr\sim\crcr}}}

\def\be{\begin{equation}}
\def\ee{\end{equation}}
\def\bea{\begin{eqnarray}}
\def\eea{\end{eqnarray}}

\def\sigv{\langle\sigma v\rangle}

\begin{document}

\hfill CERN-PH-TH/2009-025, FERMILAB-PUB-09-045-A

\title[Gamma rays from Dark Matter Annihilation in the Central Region of the Galaxy]{Gamma rays from Dark Matter Annihilation in the Central Region of the Galaxy}
\author{Pasquale Dario Serpico}
\address{Physics Division, Theory Group, CERN, CH-1211 Geneva 23, Switzerland}
\eads{serpico@cern.ch}
\author{Dan Hooper}
\address{Theoretical Astrophysics, 
Fermi National Accelerator Laboratory, Batavia, USA}
\address{Department of Astronomy and Astrophysics, 
The University of Chicago, USA}
\eads{dhooper@fnal.gov}

\begin{abstract}
In this article, we review the prospects for the Fermi satellite (formerly known as GLAST) to detect gamma rays from dark matter annihilations in the Central Region of the Milky Way, in particular on the light of the recent astrophysical observations and discoveries of Imaging Atmospheric Cherenkov Telescopes. While the existence of significant backgrounds in this part of the sky limits Fermi's discovery potential to some degree, this can be mitigated by exploiting the peculiar energy spectrum and angular
distribution of the dark matter annihilation signal relative to those of astrophysical backgrounds. \\{}\\
\noindent {\em Keywords}: Dark Matter, Gamma Rays, Galactic Center
\end{abstract}
\maketitle
\section{Introduction}\label{intro}
Despite the numerous  cosmological and astrophysical indications of the presence of non-baryonic dark matter (DM), the particle nature of this substance remains unknown. If the DM consists of weakly interacting massive particles (WIMPs), an important tool for inferring their properties could be ``indirect detection'', i.e. astrophysical observations of the annihilation (or decay) products of DM in our Galaxy or beyond.  For WIMPs with masses at or around the electroweak scale, $m_X\sim\mathcal{O}$(0.1-1) TeV,
the annihilation products are typically found at GeV-TeV energies, the domain of high-energy astrophysics. Of the different annihilation products, gamma rays and neutrinos have the important advantage of retaining directional information while not suffering energy losses. The very small cross sections of neutrinos, however, make their flux from the region of the Galactic Center very difficult to detect. On the contrary, the gamma-ray spectrum from DM annihilation or decay may be detectable with sufficient statistics, energy
resolution, and over an extended angular distribution, to provide a very distinctive set of information related to both the particle identity of the WIMP and its astrophysical distribution. A major challenge is in separating the DM signal from any astrophysical backgrounds, whose energy spectrum and angular distribution are not well known.

A major change in the prospects for DM detection has occurred in the last few
years, following the discovery of a bright astrophysical source of TeV gamma rays from the Galactic Center. As a result of this discovery, we now know that DM emission from the Galactic Center will not be detectable in a (quasi) background-free regime,
and---unless one turns the attention to other targets---the peculiar spectral shape and angular distribution of the signal must be used to extract it from this and other backgrounds.  This topic is the main subject of this review.
The remainder of this paper is structured as follows: In Sec.~\ref{instr} we briefly describe the capabilities of current high energy gamma-ray telescopes. In Sec.~\ref{signal} we
review the features of the DM annihilation signal, both in its energy (Sec.~\ref{espectra}) and angular dependence  (Sec.~\ref{angshape}). In Sec.~\ref{GC} we review the prospects for detecting or constraining DM annihilations in the Galactic Center region, while in Sec.~\ref{innerhalo} we briefly discuss some interesting aspects of the diffuse signal from the inner halo. In Sec.~\ref{conclusions} we report our conclusions.
Appendix A briefly illustrates the qualitative changes to the considerations we develop
in the main text when decaying (as opposed to annihilating) DM is considered. Appendix B 
describes a subtlety regarding the angular shape of the signal arising for a velocity-dependent 
annihilation cross section.

\section{The Instruments}\label{instr}
The ability of gamma-ray experiments to identify DM annihilation radiation from the Galactic Center region relies on the effective area, angular and energy resolution of the existing telescopes,
as well as the rejection of other (mostly hadronic) cosmic ray events contaminating the gamma-ray sample.
Since the mean free path of gamma rays is much shorter than the atmospheric slant depth, direct observations in the GeV region and above can only be done from space---which is the strategy
pursued by the LAT detector on the Fermi gamma-ray space telescope (formerly GLAST)~\cite{GLASTurl}---or indirectly by ground-based Imaging Atmospheric Cerenkov Telescopes (IACTs) such as HESS~\cite{HESSurl}, 
MAGIC~\cite{MAGICurl}, VERITAS~\cite{VERITASurl} and CANGAROO-III~\cite{CANGAROOIIIurl}.
In the latter category, the direction and energy of the primary particle hitting the atmosphere is reconstructed from the Cherenkov emission of the secondary charged particles generated in the atmospheric shower. 

These differences lead the two classes of experiments to adopt different strategies in the search for DM. Fermi-LAT is very effective in rejecting hadronic events, and continuously monitors a large fraction of the sky, but has an effective area of only $\sim 1\,$m$^2$, far smaller than that of ground-based telescopes, $\sim 10^4\,$m$^2$.  On the other hand, IACTs study small angular fields and have a lower rejection capability, but much greater overall exposure. As a consequence, diffuse gamma-ray signals are better probed by Fermi-LAT.  Any unidentified sources detected by Fermi-LAT which lack a low-energy counterpart could be potentially attributed to DM substructure. IACTs would be very effective in providing detailed follow-up observations of such sources.

In addition, the accessible energy range is very different between these two classes of experiments: $\sim$100 MeV to 300 GeV for Fermi-LAT, and above $\sim$100 GeV for ACTs. This difference makes Fermi-LAT most sensitive to DM particles lighter than a few hundreds GeV, while IACTs are better suited for TeV-scale or heavier WIMPs. On the other hand, the Fermi-LAT has poorer angular resolution than
IACTs, so it is less accurate in the localization of point-like sources. 
For both instrument classes, the search for indirect DM signatures is among the top physics priorities.
The reach of Fermi-LAT and current and future IACTs has been recently assessed in Ref.~\cite{Baltz:2008wd} and
 Ref.~\cite{Buckley:2008zc}, respectively.

\section{The Dark Matter Signal}\label{signal}

The differential flux  of gamma rays (photons per unit area, time, energy and steradian) produced in DM annihilations\footnote{See Appendix A for the case of DM decay.} is described by
\begin{equation}
\Phi_{\gamma} (E_\gamma,\Omega)= \left[ \frac{\d N_{\gamma}}{\d E_{\gamma}} (E_\gamma)\frac{\sigv}{8\pi m^2_X} \right]\int_{\rm{los}} \rho^2(\ell,\Omega)\, \d \ell,
\label{flux1}
\end{equation}
where $\sigv$ is the WIMP annihilation cross section multiplied by the relative velocity of the two WIMPs 
(averaged over the WIMP velocity distribution), $m_X$ is the mass of the WIMP, $\rho$ is the position-dependent DM density, and the integral is performed over the line-of-sight (los) in the direction of the sky, $\Omega$. The gamma-ray spectrum generated per WIMP annihilation is $\d N_{\gamma}/\d E_{\gamma}$, it has units of Energy$^{-1}$ and its integral over energy is equal to 1. If the DM is not its own antiparticle as assumed here, Eq. (\ref{flux1}) should be multiplied
further by a factor 1/2 (if $X$ and $\bar{X}$ are equally abundant).
 The factor in square brackets in Eq.~(\ref{flux1}) depends only on particle physics: in particular,
cross section, mass, and the spectrum of gamma rays produced through DM annihilations depends on the nature of the WIMP.  The integral over the line-of-sight determines instead the angular dependence of the signal and is controlled by the astrophysical distribution of DM. 
We shall discuss each of these two terms in the following, assuming that this factorization holds (see~\ref{entanglingEandAng} for a discussion of possible violations of this hypothesis).

In convenient units, Eq.~(\ref{flux1}) can be recast as: 
\begin{equation}
\frac{\Phi_{\gamma}(E_{\gamma},\Omega)}{{\rm cm}^{-2} \, {\rm s}^{-1}\, {\rm sr}^{-1}} \approx 2.8 \times 10^{-10}\,J(\Omega)\, \frac{\d N_{\gamma}}{\d E_{\gamma}}(E_\gamma)\frac{\sigv}{\rm pb}\,  \left(\frac{100\, \rm{GeV}}{m_{\rm{X}}}\right)^2.
\label{flux2}
\end{equation}
The dimensionless function $J(\Omega)$ depends only on the DM distribution in the halo and is defined by convention as~\cite{Bergstrom:1997fj}
\begin{equation}
J(\Omega) = \frac{1}{8.5 \, \rm{kpc}} \bigg(\frac{1}{0.3 \, \rm{GeV}/\rm{cm}^3}\bigg)^2 \, \int_{\rm{los}} \rho^2(\ell,\Omega) {\rm d} \ell\,.
\label{jpsi}
\end{equation}
For typical halo models (see Sec. 3.2) this is a function strongly peaked towards the Galactic Center (for an illustration, see for example Fig. A1). 

The benchmark value for the cross-section, $\sigv \approx $1 picobarn (or in cgs units $\sim 3 \times 10^{-26} \,\rm{cm}^3/\rm{s}$), is motivated by the fact that a WIMP annihilating with such a cross section during the freeze-out epoch will be generated as a thermal relic with a density similar to the measured DM abundance (for a review, see Ref.~\cite{Bertone:2004pz}). WIMPs constituting the {\it cold} DM annihilate in the non-relativistic limit. If annihilations take place largely through $S$-wave processes, then the annihilation cross section of WIMPs in the Galactic halo (i.e. in the low velocity limit) will also be approximately equal to this value, which justifies the benchmark value used in Eq.~(\ref{flux2}). Yet, it is important to 
stress that much lower signals are possible (e.g. if $P-$wave annihilation dominates the
freeze-out process), as well as  significantly enhanced ones in models where the DM is non-thermally produced in the early universe (see e.g. Ref.~\cite{Moroi:1999zb}). Some further considerations
and refs. can be found in~\ref{entanglingEandAng}. 

It is also worth noting that for S-wave annihilating thermal  relics the indirect detection signal has two advantages compared to direct detection via nuclear recoils in underground detectors:
i) it is proportional to a relatively large annihilation cross section; ii) it is less dependent from the particle physics details. For example, even under the (possibly unrealistic) assumption that DM annihilates mostly into quarks of the first generation, the natural expectation value for DM-nucleon elastic cross section is at the level of $(m_N/m_X)^2\times 1\,$pb$\approx 10^{-40}\,(m_X/0.1\,{\rm TeV})^2\,$cm$^2$. 
For DM annihilating predominantly into heavier particles, a further suppression is expected.

\subsection{Energy Spectra}\label{espectra}
\begin{figure}
\begin{center}\resizebox{7.0cm}{!}{\includegraphics{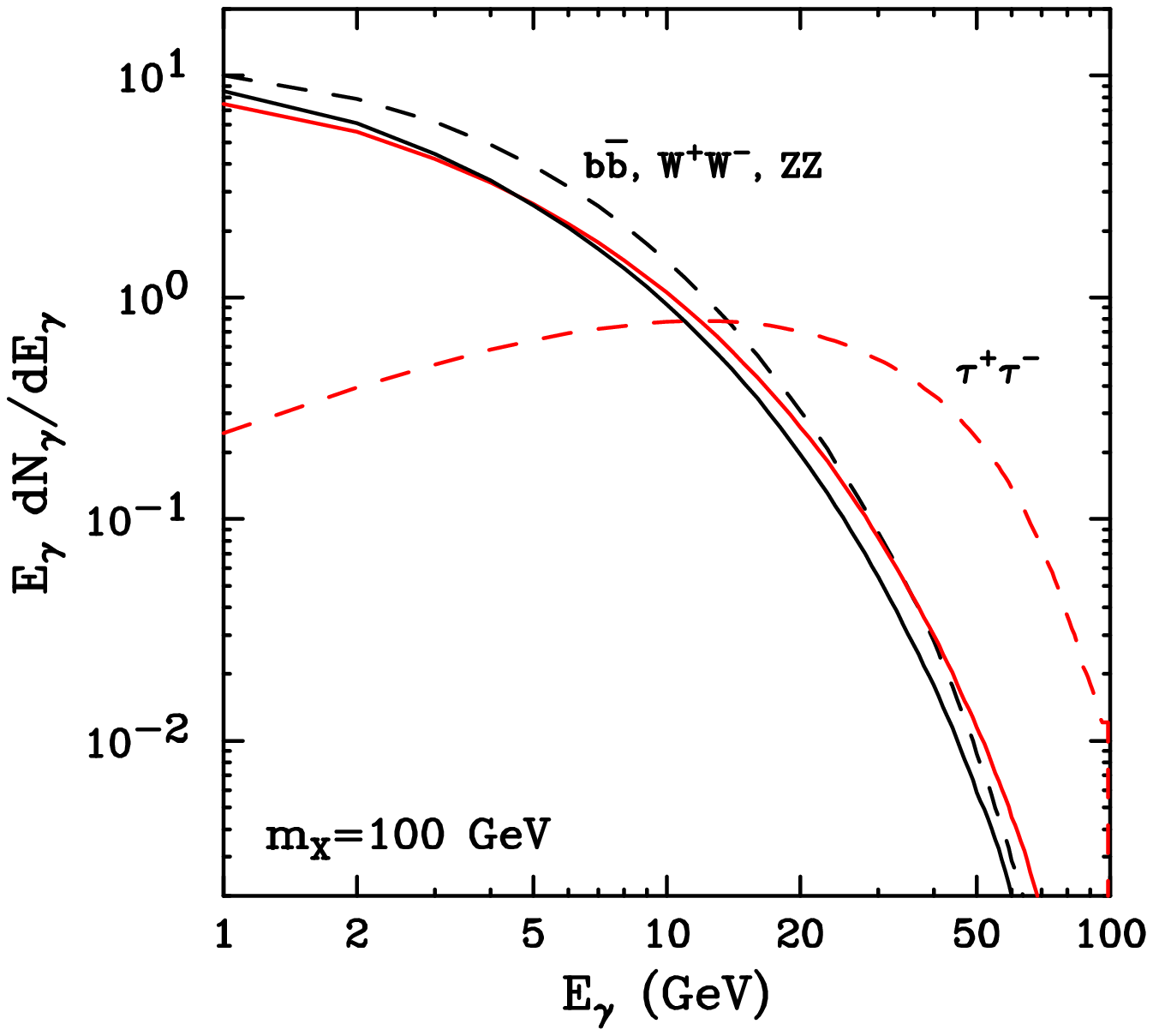}} 
\hspace{0.1cm}
\resizebox{7.0cm}{!}{\includegraphics{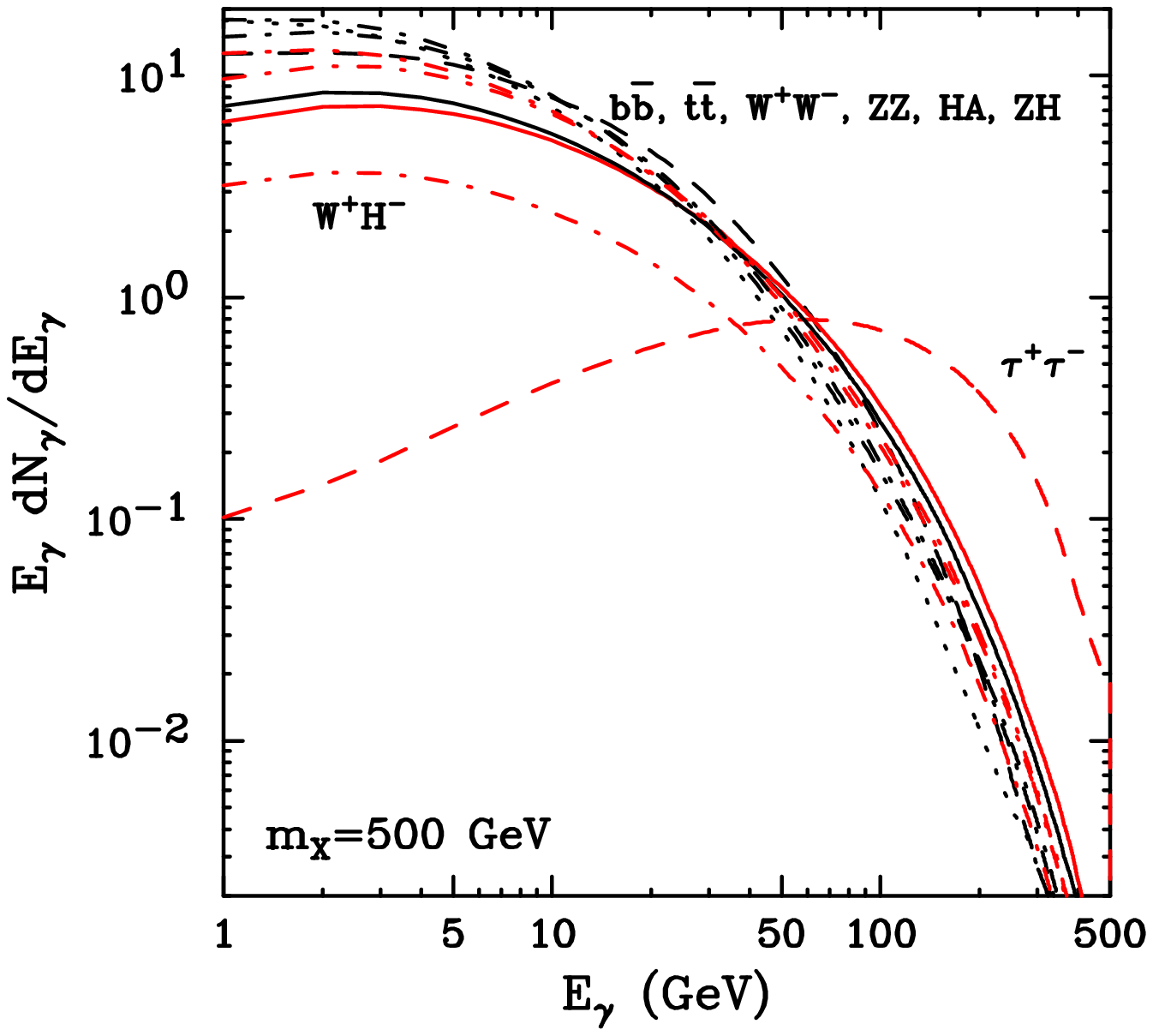}}
\caption{The gamma-ray spectrum per annihilation for a 100 GeV (left) and 500 GeV (right) WIMP. Each curve denotes the result for a different dominant annihilation mode. From Ref.~\cite{Dodelson:2007gd}.}
\end{center}
\label{spectra}
\end{figure}

The gamma-ray spectrum from DM annihilation originates from several different contributions. Typically,  the most abundant source of photons is the hadronization and/or decays of unstable particles. For example, neutralinos in the Minimal Supersymmetric Standard Model (MSSM), dominantly annihilate to final states consisting of heavy fermions $b \bar{b}$, $t \bar{t}$, $\tau^+ \tau^-$ (i.e. bottom quarks, top quarks and tau leptons, respectively) or bosons $ZZ$, $W^+ W^-$, $HA$, $hA$, $ZH$,
$Zh$, $ZA$, $W^{\pm} H^{\pm}$, where $W^{\pm},Z$ are the gauge bosons mediating the weak interactions and $H$, $h$, $A$ and $H^{\pm}$ are the Higgs bosons of the MSSM~\cite{jungman}.
With the exception of the $\tau^{+} \tau^-$ channel, each of these annihilation modes result in a very similar spectrum of gamma rays, dominated eventually
by the decay of mesons, especially $\pi^0$, generated in the cascade. In Fig.~1 we show the predicted gamma-ray spectrum per annihilation, for several possible WIMP annihilation modes.  The harder spectrum here shown from annihilation into the $\tau^{+}\tau^{-}$ channel
is not typical of SUSY models, although it might be a distinctive feature of DM annihilating  dominantly to charged leptons pairs; this  arises, for example, in Kaluza-Klein DM models (for a review of particle DM models, see e.g.~\cite{Bergstrom:2009ib}).

Although these secondary photons provide the dominant emission, other important channels can exist at the one-loop level. A particularly striking signature would be the mono-energetic photons resulting from final states such as $\gamma \gamma$, $\gamma Z$ or $\gamma h$ (see Ref.~\cite{lines} for a discussion of these processes within the context of supersymmetry). Unfortunately, such processes are expected to produce far fewer events than continuum emission and in typical models can not easily be detected (for a counterexample, see Ref.~\cite{Gustafsson:2007pc}).  Bremsstrahlung (with an additional photon appearing in the final state) is automatically present when annihilations produce charged final states, and can dominate the high energy region of the spectrum when those charged annihilation products are much lighter than the WIMPs. This is particularly important when the tree-level processes to a pair of light fermions are disfavored by ``selections rules'', but no suppression is present for  three body final states~\cite{Birkedal:2005ep,Bringmann:2007nk}. For more details, see the review by L.~Bergstrom in this issue~\cite{Bergstrom:2009ib}.  Note that  gamma-ray emission from DM annihilating into standard model particles is unavoidable, even in the most extreme case when only neutrino final states are allowed at tree level,
due to $W$ and $Z$-strahlung~\cite{Kachelriess:2007aj,Bell:2008ey}. 

\subsection{Angular Shape Of The Signal}\label{angshape}

The gamma-ray spectrum and angular distribution predicted by Eq.~(\ref{flux1}) is rather general (see, however, Appendix B), and could be applied to the case of a smoothly distributed Galactic halo, or alternative targets such as dwarf spheriodal galaxies, microhalos/clumps, density spikes around intermediate mass black holes, or the integrated extragalactic diffuse background (described in other contributions to this special issue). While some of these targets have interesting observational prospects, the intensity of these signals is strongly dependent on often unknown cosmological and astrophysical properties, such as the quantity of small-scale structures in DM halos or the population of intermediate mass black holes in the Milky Way. In this article, we will focus on the smooth Galactic halo.

Naively, by inputing the DM density profile inferred from kinematical observations of the Milky Way into Eq.~(\ref{flux1}), one could obtain an approximate lower bound on the gamma-ray flux from DM annihilation. In principle, this would enable us to translate the observations from ground or space-based gamma-ray telescopes into constraints on the particle physics properties of the WIMP (mass, annihilation cross section and dominant modes). Unfortunately, even the average, smooth distribution
of the DM particles in our Galaxy is not well known, especially in the volume within the solar circle. This is due to the fact that the baryonic material dominates the gravitational potential in the inner Galaxy, and lacking a detailed knowledge of its distribution, a reconstruction of the DM distribution ``by subtraction'' is unfeasible. It is not surprising, then, that very different profiles have been claimed to fit the observations (for example, compare the results of Ref.~\cite{Binney:2001wu} and Ref.~\cite{Klypin:2001xu}). 

On general grounds, a class of spherically symmetric, smooth halo distributions can be used to approximately fit both the observed rotation curves of galaxies and the results of numerical simulations of DM halos. A non-trivial angular dependence of the gamma-ray signal results from the off-center position
of the Sun within the halo (see e.g. Ref.~\cite{Hooper:2007be} and references therein for a discussion of sub-leading effects determining the angular distribution of the signal).
The function $J$ introduced in Eq.~(\ref{flux2}) and Eq.~(\ref{jpsi}) then depends only on the angle $\theta$ between the observed direction of the sky and the Galactic Center  or, in terms of galactic latitude $b$ and
longitude $l$,  only on $\cos\theta=\cos b\cos l$. The radial variable, $r$, can be expressed in terms
of the relevant quantities $\{\ell,\theta\}$ as
\begin{equation}
r(\ell,\theta)=\sqrt{r_\odot^2+\ell^2-2\,r_\odot\,\ell\cos\theta}\,, \label{rspsi}
\end{equation}
where $r_\odot\approx 8.33\pm 0.35\,$kpc~\cite{Gillessen:2008qv}
is the distance of the Solar System from the Galactic Center.
Typically, one considers a distribution of the form
\begin{equation}
\rho(r)=\left(\frac{r_s}{r}\right)^\gamma
\frac{\rho_0}{[1+(r/r_s)^\alpha]^{(\beta-\gamma)/\alpha}},\label{prof}
\end{equation}
where $\rho_0$ is a normalization constant and $r_s$ is a characteristic  radius
below which the profile scales as $r^{-\gamma}$. 
A very well known, universal profile of this class fit to DM-only ({\it i.e.} neglecting baryons) N-body simulations has been proposed by Navarro, Frenk and White
(NFW)~\cite{NFW}, corresponding to the choice $\{\alpha,\beta,\gamma\}=\{1.0,3,1.0\}$. Steeper or softer profiles have also been extensively discussed, such as that proposed by Moore et al.~\cite{Moore} and 
Kravtsov et al.~\cite{Kravtsov:1997dp}, respectively.  While simulations (and data as well) typically agree on the shape of profile in the outskirts of the halos, a disagreement clearly exists concerning the inner slope, $\gamma$.  More recent simulations~\cite{Power:2002sw,Navarro:2003ew,Reed:2003hp,Merritt:2005xc,Navarro:2008kc} suggest that halo density profiles are better represented by a function with a
continuously-varying slope, as the one proposed by Einasto~\cite{Einasto}
\begin{equation}
\rho(r)=\rho_{-2}\,e^{-\frac{2}{\alpha}\left[\left(\frac{r}{r_{-2}}\right)^\alpha-1\right]}\,,\label{einasto}
\end{equation}
with clear hints for non-universality in the form of halo-to-halo variations in the quantity $\alpha$.  The quantity $r_{-2}\simeq 25\,$kpc denotes the radius at which the logarithmic slope of the profile, $\d \log\rho/\d \log r$, assumes the value $-2$; the other free parameters are $\alpha$ and the overall normalization, here chosen as the density $\rho_{-2}\equiv \rho(r_{-2})$. At $r \gsim 1$ kpc, these newly proposed fitting formulae provide only marginal improvement with respect to the more
traditional ones. At smaller radii, however, these recent results lead us to expect the inner slope of DM halos to be shallower that that predicted by Moore {\it et al.}, and probably shallower than NFW, as well.

\begin{table}[!tbh]\label{TableII}
\begin{center}\begin{tabular}{c|ccccc}
\hline\hline
Model    & $\alpha$ & $\beta$ & $\gamma$ &$\rho_\odot$ & $r_s$ \\
  & &  &  &[GeV$\,$cm$^{-3}$]& [kpc]\\
\hline\hline
Moore {\it et al.}    & 1.5 & 3 & 1.5 &0.27 & 28\\
NFW          & 1.0 & 3 & 1.0 & 0.30 & 20\\
Kravtsov   & 2.0 & 3 & 0.4 & 0.37 & 10\\
\hline
\end{tabular}\caption{Parameters describing some common halo profiles of the form described by Eq.~(\ref{prof}), where $\rho_\odot$ is the DM density at the Solar distance from the GC. See text for more details.}

\end{center}
\end{table}

There are a number of astrophysical processes that may potentially modify the DM distribution, none of which are taken into account in the above-mentioned profiles. It is very difficult to reliably account for these effects in simulations, and only a few results are available (see e.g. Refs.~\cite{Gnedin04,Gustafsson:2006gr,Read:2009iv}). 
Qualitatively, since baryons can cool and contract, one expects them to steepen the gravitational potential in the central regions galaxies and, as a result, enhance the DM density~\cite{ac}. On the other hand, other feedback or frictional effects have been proposed that could reduce the DM density in the inner halos and bring the prediction closer to observations (for a recent review, see Ref.~\cite{Sellwood:2008bd}). For some galaxies, it has actually been argued that flat-cored profiles fit the observations better than cusped profiles.

Even greater uncertainties exist concerning the DM distribution at the very center of the Milky Way, in the region immediately surrounding the central supermassive black hole. Adiabatic accretion may lead to the formation of a spike in the DM distribution, resulting in a very high DM annihilation rate in the innermost parsecs of the galaxy~\cite{spike}. Mergers as well as scattering on the dense stellar cusp around the central black hole may potentially destroy density enhancements, however. In general, these effects only affect regions too close to the Galactic Center to be resolved angularly by present detectors, leaving only the energy spectrum to be used for separating the DM signal from the background. Yet, such a spike might lead to a measurable gamma-ray flux from the innermost angular bin, even in presence of
relatively large astrophysical backgrounds. For more details, we direct the reader to the references cited in Ref.~\cite{Bertone:2004pz} or Ref.~\cite{Fornasa:2007nr}.

Given these considerably uncertainties, we have chosen to use three illustrative choices for the DM halo profiles: the Moore {\it et al.}, NFW, and Kravtsov profiles (with the parameters given in Table I). In each case, as in Ref.~\cite{Yuksel:2007ac}, the normalization has been chosen so that the mass contained within
the solar circle provides the appropriate DM contribution to the local rotational curves.
In all of the cases we have discussed, even for the most conservative cored profile, the DM signal peaks at the Galactic Center. In presence of an isotropic (or even vanishing) astrophysical background, the Galactic Center region thus becomes the natural location to look for a DM signal. Unfortunately, there is also a higher
background density as we look toward the inner region of our Galaxy. Still, as a first
step, one can consider the detection prospects for this region, a task we address in Sec.~\ref{GC}, 
before turning to more general arguments regarding the optimal regions of the sky for DM searches in Sec.~\ref{innerhalo}.

\section{The Galactic Center}\label{GC}
\begin{figure}
\begin{center}
\includegraphics[width=.6\textwidth]{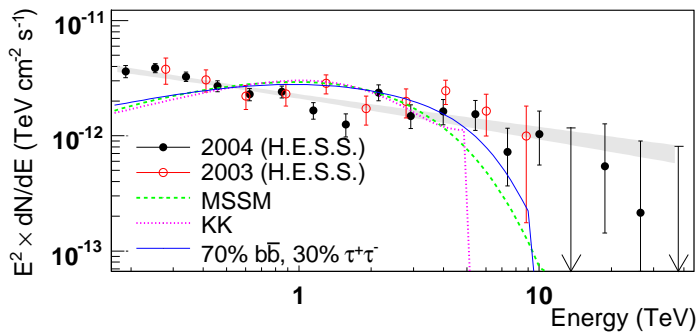}
\caption{\label{fig:edis} 
Spectral energy density of the Galactic Center source as measured by
HESS in 2004 (full points) and 2003~\cite{hess} (open points).
Upper limits are 95\% CL.
The shaded area shows the power-law fit $\mathrm{d}N/\mathrm{d}E \sim E^{-\Gamma}$, with $\Gamma=2.25\pm0.04\,$(stat.)$\pm 0.10\,$(syst.).
The dashed line illustrates typical spectra of phenomenological MSSM DM annihilation 
for best fit neutralino masses of 14~TeV.
The dotted line shows the distribution predicted for 
Kaluza-Klein DM with a mass of 5~TeV. 
The solid line gives the spectrum of a 10~TeV DM particle
annihilating into $\tau^+\tau^-$ (30\%) and $b\bar{b}$ (70\%).  From Ref.~\cite{Aharonian:2006wh}.}
\end{center}
\end{figure}

%
The Galactic Center has long been considered to be among the most
promising targets for the detection of DM annihilation, particularly if the halo profile of the Milky Way is cusped in its inner volume~\cite{Bergstrom:1997fj,gchist}. This has been complicated, however, by the recent discovery of astrophysical gamma-ray sources from the Galactic Center. Following an earlier claim by the WHIPPLE IACT~\cite{whipple}, very high-energy gamma rays from the Galactic Center have been detected by HESS~\cite{hess}, MAGIC~\cite{magic}, and CANGAROO-II~\cite{cangaroo}. This source is consistent with point-like emission and is located at $l = 359^\circ 56^\prime 41.1^{\prime\prime}\pm 6.4^{\prime\prime}$ (stat), 
$b = -0^\circ2^{\prime}39.2^{\prime\prime}\pm 5.9^{\prime\prime}$ (stat) with a systematic pointing error of
28$^{\prime\prime}$ \cite{van Eldik:2007yi}, coincident with the position of Sgr A$^{\star}$, the black hole constituting 
the dynamical center of the Milky Way. The spectrum of this source is well described by a power-law
with a spectral index of $\alpha=2.25 \pm 0.04 (\rm{stat}) \pm 0.10 (\rm{syst})$ 
over the range of approximately 160 GeV to 20 TeV. 
Although speculations were initially made that this source could 
be the product of annihilations of very heavy ($\sim$10 to 50 TeV) DM particles~\cite{actdark}, this interpretation is disfavored by the power-law form of the observed spectrum and the wide energy
range over which it extends (see Fig.~\ref{fig:edis}). 
The source of these gamma rays is, instead, likely an astrophysical 
accelerator associated with our Galaxy's central supermassive black hole~\cite{hessastro}.
In recent analyses, this source has been treated as a background for DM searches~\cite{Zaharijas:2006qb,Hooper:2007gi}.  Given the presence of this background, the prospects for detecting DM annihilation products from the Galactic Center appear considerably less promising than they had a few years ago. 

The main challenge involved in DM searches with Fermi will be to distinguish the signal from this and other backgrounds.  This is made particularly difficult by our ignorance regarding the nature of these backgrounds. The Galactic Center is indeed a complex region of the sky at all wavelengths, the gamma-ray window being no exception~\cite{vanEldik:2008ye}.

An attempt to use angular and spectral information to separate DM annihilation products from these backgrounds was performed in Ref.~\cite{Dodelson:2007gd}, studying a $2^\circ\times 2^\circ$ region arond the Galactic Center (see also the analysis of the Fermi DM team in Ref.~\cite{Baltz:2008wd}). This study considered two known point-source backgrounds: a yet unidentified source detected by EGRET approximately 0.2$^{\circ}$ away from the dynamical center of our galaxy~\cite{dingus,pohl}, and the IACT source discussed above. In addtion, a diffuse spectrum with a free power-law index was also included. 
More in detail, we describe the spectrum of the source revealed by IACTs at the Galactic Center as a power-law given by:
\begin{equation}
\Phi^{\rm ACT} = 1.0 \times 10^{-8} \left({ {E_{\gamma}}\over{{\rm GeV}}}\right)^{-2.25} {\rm GeV}^{-1} \, {\rm cm}^{-2} \, {\rm s}^{-1}.
\end{equation}

The flux from the EGRET source slightly off-set is instead modeled  as:
\begin{equation}
\Phi^{\rm EG} = 2.2 \times 10^{-7} \,\left({ {E_{\gamma}}\over{{\rm GeV}}}\right)^{-2.2} e^{-\frac{E_{\gamma}}{30 \, {\rm GeV}}}\, {\rm GeV}^{-1} \, {\rm cm}^{-2} \, {\rm s}^{-1}  \,,
\end{equation}
where the exact value of the cutoff energy (here 30 GeV) is somewhat arbitrary, but reflects the fact that this source has not been
observed yet by IACTs at $E\gsim 100\,$GeV. Finally, we allow for a diffuse/unresolved flux with spectrum
\begin{equation}
\Phi^{\rm diff}(A,\alpha) = A \left({ {E_{\gamma}}\over{{\rm GeV}}}\right)^{-\alpha} {\rm GeV}^{-1} \, {\rm cm}^{-2} \, {\rm s}^{-1} \, {\rm sr}^{-1}\,,
\end{equation}
where $\alpha$ is allowed to vary between 1.5 and 3.0. We adopt an overall normalization, $A$, such that the integrated flux
of the diffuse background between 1 GeV and 300 GeV in a $2^\circ\times 2^\circ$ field of view around the Galactic Center
is equal to $10^{-4} \, {\rm cm}^{-2} \, {\rm s}^{-1} \, {\rm sr}^{-1}$. We do not, however, assume that this normalization is known in our analysis, leaving open the possibility that some of the diffuse gamma rays observed are the product of dark matter annihilations.

\begin{figure}
\begin{center}
\resizebox{7.5cm}{!}{\includegraphics{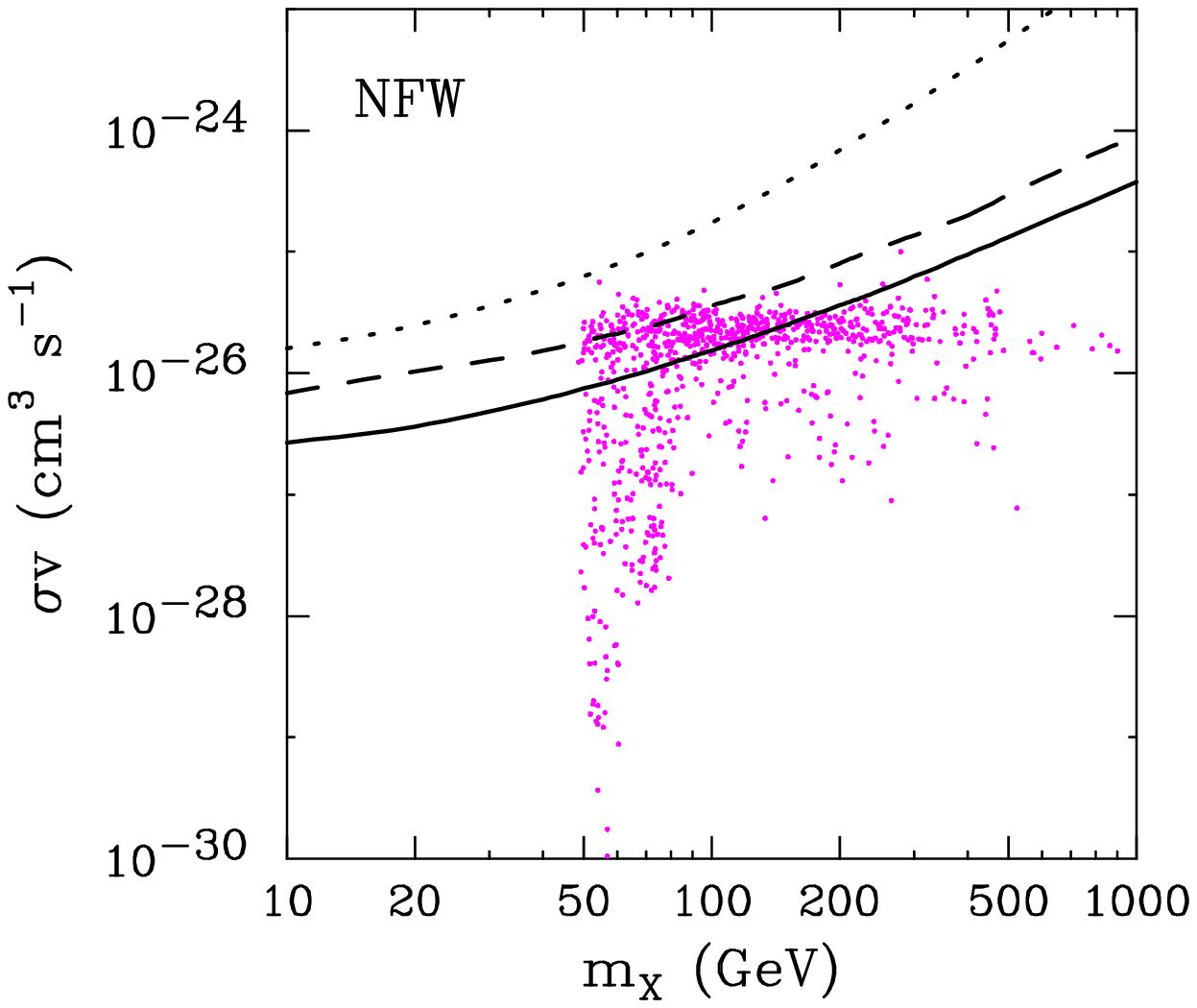}}
\resizebox{7.5cm}{!}{\includegraphics{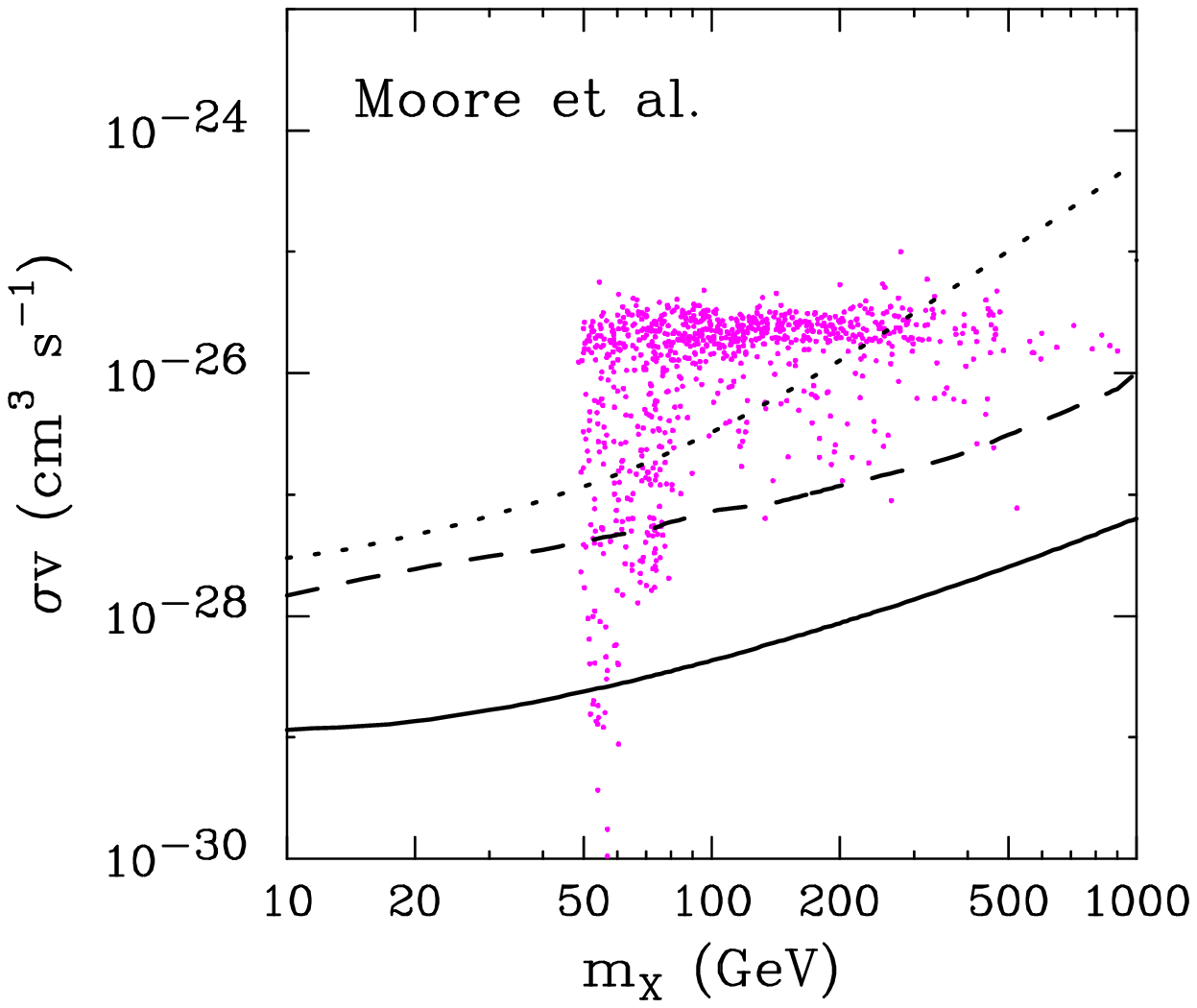}}
\caption{The projected exclusion limits at 95\% confidence level from Fermi-LAT (after ten years) on the WIMP annihilation cross section,  as a function of the WIMP mass. The region above the dotted line is already excluded by EGRET~\cite{dingus}. The solid and dashed lines show the projections
for Fermi-LAT  for an assumed isotropic diffuse background and the limiting case where the astrophysical background has exactly the same angular distribution of the DM signal, respectively.  In the left and right frames, the NFW and Moore {\it et al}.~halo profiles have been adopted, respectively.  Also shown are points representing a random scan of supersymmetric models. From Ref.~\cite{Dodelson:2007gd}.}
\label{limitnfw}
\end{center}
\end{figure}

A multi-parameter $\chi^2$ analysis of the simulated sky against models including a contribution from DM annihilation radiation yields  the projected exclusion limits at the 95\% confidence level shown in Fig.~\ref{limitnfw} (for ten years of collection time by Fermi-LAT).  The simulated sky used to produce this figure contains the two resolved sources and the diffuse background but no dark matter, and to derive the exclusion limits  we compare the signal thus obtained to a model which includes both the backgrounds and a signal from dark matter. The left and right panels refer to the cases of an NFW profile and Moore {\it et al.} profile (replaced by a flat core within $10^{-2}$ pc of the Galactic Center to avoid a divergency), respectively. The WIMP is always assumed to annihilate dominantly to $b \bar{b}$. This assumption is not particularly restrictive, since for many annihilation modes the spectrum would look very similar (see Fig.~1). Apart for some cases, as possibly Kaluza Klein DM, the
gamma-ray spectrum alone does not help much in discriminating among several DM candidates, see e.g.~\cite{Hooper:2006xe} for details. 
\begin{figure}
\begin{center}
\resizebox{7.5cm}{!}{\includegraphics{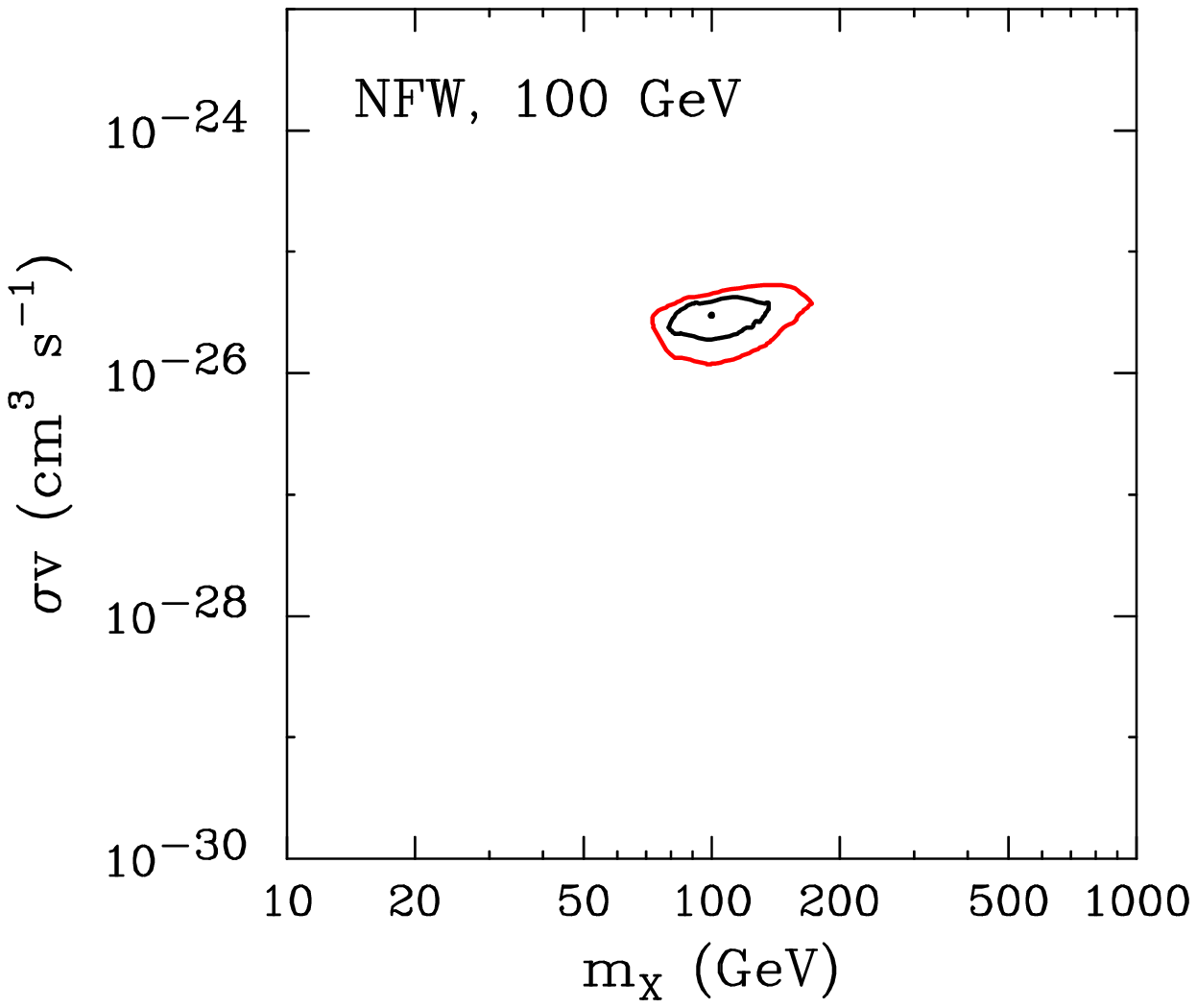}}
\resizebox{7.5cm}{!}{\includegraphics{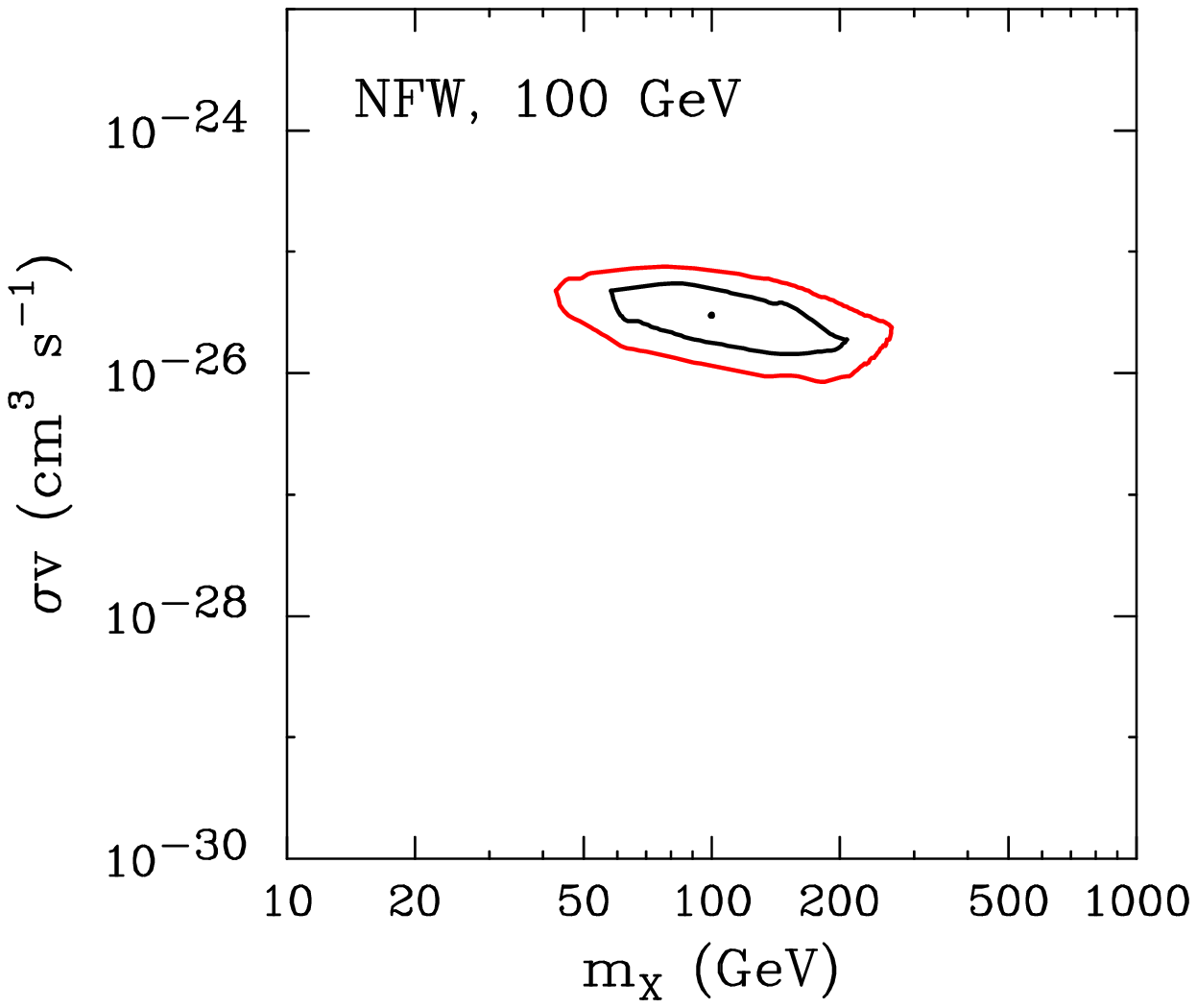}}
\caption{The ability of Fermi-LAT to measure the annihilation cross section and mass of dark matter after ten years of observation. A benchmark scenario with $m_X=100$ GeV, $\sigv = 3 \times 10^{-26}$ cm$^3$/s and an NFW halo profile has been used. The inner and outer contours in each frame represent the 2 and 3$\,\sigma$ regions, respectively. In the left frame, the halo profile shape was treated as if it is known in advance. In the right frame, a marginalization over the inner slope of the profile was performed. From Ref.~\cite{Dodelson:2007gd}.}
\label{ellipsenfw}
\end{center}
\end{figure}

To give a feeling for the numbers involved, the integral of the function $J$ over the considered region leads to a factor $\sim 1.3$ for the NFW model; in this model, for the benchmark 100 GeV WIMP with $\sigv \sim 3 \times 10^{-26}$ cm$^3$/s one
would expect less than 3000 events from DM above 1 GeV, to be compared with more than 2.3$\times 10^5$ background photons. The background is almost evenly split between the EGRET point source and the diffuse flux; the IACT source contributes less than 7000 events, but it is located at the GC and, differently from the EGRET source, is not cut-off at high energy, so it is important to include it especially for high WIMP masses. 
The solid line in each frame represents the limit found if the diffuse background is assumed to be distributed isotropically, while the dashed line represents the conservative limit obtained if the diffuse background has the same angular distribution as the DM signal ({\it i.e.} the case in which angular information is not useful in disentangling the signal from the diffuse background). For values of $\sigv$ below the corresponding lines, a pure background model  is expected to be consistent with the data.
The fact that the limits are significantly stronger in the uniform background case is the manifestation of the
improved sensitivity which can be achieved by an analysis including both energy and angular information.
 For comparison,  in Fig.~\ref{limitnfw}  it is also shown the region already excluded by EGRET~\cite{dingus} (above the dotted line) and the mass and cross section of neutralino models found in a random scan over supersymmetric parameters, as calculated using DarkSUSY~\cite{darksusy}. As expected, many of the models cluster around $\sigv \sim 3 \times 10^{-26}$ cm$^3$/s, the value required of a thermal relic annihilating via an $S$-wave amplitude. Each point shown represents a model which respects all direct collider constraint and generates a thermal DM abundance consistent with the observed DM density. In the scan  the SUSY parameters varied were $M_2$, $|\mu|$ and $m_{\tilde{q}}$ up to 2 TeV, $m_A$ and $m_{\tilde{l}}$ up to 1 TeV and $\tan \beta$ up to 60. Also,  the gaugino masses were assumed to evolve to a single unified scale, such that $M_1 \approx 0.5 M_2$, $M_3 \approx 2.7 M_2$.


Should gamma rays be identified as having been produced in DM annihilations, such observations could then be used to measure the characteristics of the DM particle, including its mass, annihilation cross section and spatial distribution. Such determinations are an important step toward identifying the particle nature of DM. A calculation similar to the one leading to the
results of Fig.~\ref{limitnfw} can be performed, this time including in the template  a contribution from a fiducial model of DM annihilation, and asking the accuracy by which a reconstruction of its
input properties is possible. In Fig.~\ref{ellipsenfw}, the ability of Fermi-LAT to determine the WIMP mass and annihilation cross section for WIMPs distributed with an NFW halo profile is illustrated, for the case of an annihilation cross section of $3 \times 10^{-26} \,\rm{cm}^3/\rm{s}$ and a mass of 100 GeV. 
In each frame the projected 2 and 3$\,\sigma$ constraints on the input parameters are reported, assuming an isotropic diffuse background (in addition to background point sources). In the left frame, we treat the shape of the halo profile (NFW) as if it is known in advance. Of course, this is not a realistic assumption, and a less accurate determination of the WIMP mass must be expected in a more realistic treatment. In the right panel of Fig.~\ref{ellipsenfw} we report the results obtained by marginalizing over the inner slope of the halo profile, $\gamma$. In the absence of an assumption on the inner halo slope,
the constraint on the DM mass worsens by a factor $\sim $2.

\begin{figure}
\begin{center}
\resizebox{7.5cm}{!}{\includegraphics{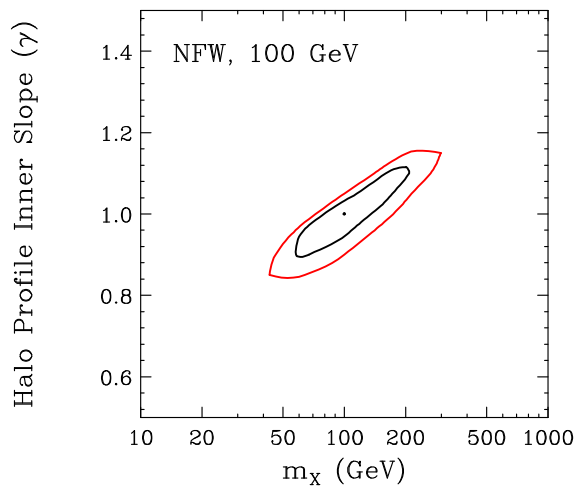}}
\caption{The ability of Fermi-LAT to measure the inner slope of the halo profile and the mass of dark matter particle (marginalizing over the annihilation cross section) after ten years of observation. Here,  a benchmark scenario with $m_X=100$ GeV, $\sigv = 3 \times 10^{-26}$ cm$^3$/s and an NFW halo profile has been used. The inner and outer contours represent the 2 and 3$\,\sigma$ regions, respectively. From Ref.~\cite{Dodelson:2007gd}.}
\label{ellipseslope}
\end{center}
\end{figure}

If the spectrum and angular distribution of gamma rays from DM annihilations in the Galactic Center region are sufficiently well measured, it will also be possible to measure the underlying DM distribution. Fig.~\ref{ellipseslope} shows  the results from the right frame of Fig.~\ref{ellipsenfw} in the $\{m_X,\,\gamma\}$ plane, marginalized over the annihilation cross section. In the above benchmark model, the inner slope of the halo profile can be determined at approximately the $\sim 10\%$ level.

A few caveats are in order: We are possibly oversimplifying the spectral shape of the background,
since we  are extrapolating the known point-like source properties from lower and higher energies. Also, the bounds are slightly optimistic, in the sense that we are considering the longest plausible lifetime of Fermi (ten years) and that further effects degrading the angular resolution
or accounting for dead-time may loosen the constraints by up to a factor$\sim\mathcal{O}$(2).
Finally, we have included only the HESS and EGRET sources (in addition to the diffuse background) in our analysis, but it is likely that other astrophysical point sources with different spectra will be discovered by Fermi-LAT. The above estimates should be quite realistic as long as a limited number of discrete sources will ``contaminate'' the inner Galactic Center angular bins, so that they could be removed effectively by the Fermi team. On the other hand, it is worth taking a lesson from the ``sudden'' discovery of an astrophysical source just at the Galactic Center:
In the worst case, the Galactic center region might so crowded with (yet unknown) sources that the 
separation of background and signal might be degraded with respect to expectations based on present knowledge.
Still, it is worth mentioning that the bulk of the statistical significance of the dark matter annihilation signal typically does not come from the inner $0.1^{\circ}$ around the Galactic Center (where the IACT source dominates), rather from the surrounding angular region. It is then interesting to see what are the perspectives to detect a DM signal from a more extended region, which we address in the following.

\section{The Inner Halo}\label{innerhalo}

\begin{figure}[!htb]
\centering
\epsfig{figure=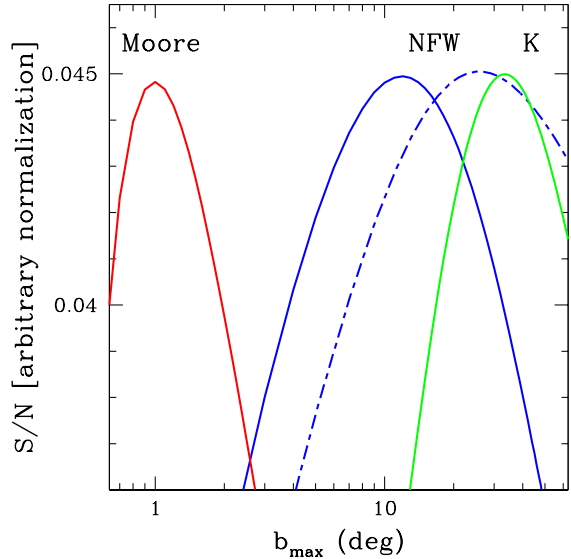,width =0.5\columnwidth,angle=0}
 \caption{\label{Fig2} 
The relative signal-to-noise as a function of the window size, $b_{\rm max}$, for a region $0.4^\circ<|b|<b_{\rm max}$, $0^\circ<|l|<l_{\rm max}=b_{\rm max}$ for the
continuum spectrum of a 100 GeV neutralino annihilating into $W^{+}W^{-}$ (dot-dashed line,
NFW profile) and for the gamma-line emission for the benchmark ``model I'' in the inert Higgs doublet scenario (solid lines). From left to right, the solid lines refer to the Moore, NFW and Kravtsov profiles, see Table I. From Ref.~\cite{Serpico:2008ga}.}
\end{figure}

The emission of radiation per unit solid angle from DM annihilation is expected to
be maximized at the Galactic Center. Yet, geometric factors and the presence of point-like
and diffuse backgrounds make the choice of the optimal window size a non-trivial problem~\cite{Baltz:2008wd,Buckley:2008zc,Bergstrom:1997fj,Stoehr:2003hf,Evans:2003sc,Serpico:2008ga}. Unless the DM halo is very cuspy towards the Galactic Center (say, cuspier than NFW), the optimal strategy for DM searches is never to focus on the inner sub-degree around the Galactic Center. Rather, a window size up to $\sim 50^\circ$ or more is preferred. The optimal shape of the window depends on the angular distribution of the signal and backgrounds, but also on the details of the analysis (like energy cuts, astrophysical foreground removal, etc.). It could be optimized from morphological studies of the low-energy emission measured by Fermi, but a circular annulus around the Galactic Center or a  ``rectangular'' window in Galactic coordinates---with an
inner rectangular mask--- generally work fairly well. Here we illustrate this issue within a simplified model: Based on EGRET data, in the diffuse background one can identify an isotropic, extragalactic component $\propto E^{-2.1}$ and a Galactic component scaling as $\sim E^{-2.7}$, which is however dominant at GeV energies and peaks towards the Galactic Plane (see \cite{Bergstrom:1997fj,Serpico:2008ga} for details).
The energy spectrum of the DM signal thus mostly enters the game in determining which one of the
two spectra (of different angular shape) dominates the background: the harder the DM spectrum, the closer the background is to an isotropic emission, the smaller the optimal angular window. Still, even in the  quite extreme\footnote{Even for  heavy dark matter particles of TeV mass,  the continuum photon spectrum peaks at or below 30 GeV, explaining in which sense the line emission considered here is ``extreme''.} case of $70\,$GeV monochromatic lines, the optimal angular window is very extended. This is illustrated in Fig.~\ref{Fig2} , where we plot the relative signal-to-noise ($S/N$) as a function of the maximum galactic latitude, $b_{\rm max}$, for a Galactic region $0.4^\circ<|b|<b_{\rm max}$, $0^\circ<|l|<l_{\rm max}=b_{\rm max}$.  
The solid lines refer to an hypothetical gamma-line emission for the benchmark ``model I'' in the inert Higgs doublet scenario (see Ref.~\cite{Gustafsson:2007pc} for details) for three different halo profiles. The dot-dashed line refers to the continuum spectrum of a 100 GeV neutralino annihilating into $W^{+}W^{-}$ with an NFW distribution, where indeed we see that the optimal window is even more extended than for the line signal: this is due to the fact that the background at the typical energies of the continuum DM photons, $ E \lsim m_X/10$, is mostly dominated by the Galactic unresolved background, which peaks towards the Galactic Plane/Center. 
To give a feeling of the quantitative difference with respect to the GC case treated in the previous section, let us note that for our usual benchmark model ($m_X=100\,$GeV, $\langle\sigma v\rangle=1\,$pb with $b\bar{b}$ final state), a ``naive'' count statistics above 1 GeV
(including all backgrounds) would lead to a $S/N~\sim 6$ for the $2^\circ\times 2^\circ$ window considered in Sec. 4, while for the region considered above and $b_{\rm max}=l_{\rm max}=25^\circ$
one would find almost 40000 events from DM vs. about seven millions background events, resulting
in a $S/N>14$. Of course, this analysis does not take into account angular and energy cuts
that (as shown for the GC case above)  do improve the diagnostic power, as well as other possible systematics. Yet, the importance of this diffuse signal cannot be underestimated (and
it is actually confirmed by other analyses, see Sec. 4.2 in Ref. [6]), especially in the case 
no signal is revealed from the GC, which may be due to a inner halo profile more cored than NFW.
The extended signal has indeed a milder dependence on the profile. Also, from Fig.~\ref{Fig2} it
is clear that---at least for the simple $S/N$ estimator---it is only important  to adopt the right angular cut within a factor of two or so in order not to degrade the sensitivity by more than$\sim 20\%$. But ``blindly'' focusing on a too narrow window (degree scale) around the GC might be overly penalizing; especially for relatively cored profiles, the loss in sensitivity may reach a factor of two or more.

Besides improving the prospects for detection, other advantages in focusing on an extended region around the Galactic Center include:
\begin{itemize}
\item  enabling an empirical determination of the dark matter profile slope outside of the region dominated by the gravitational potential of the central supermassive black hole. 
\item obtaining independent evidence that an ``excess'' signal with respect to backgrounds is the product of DM annihilation, by comparing the emission from many angular regions. In contrast, if the spectrum is found to vary with location, it is most likely the product of astrophysical backgrounds.
\item  constraining the quantity of dark matter substructure in the halo, by observing the angular distribution of the emission. The shape of the smooth, unresolved inner profile~\cite{Hooper:2007be} could be studied, in addition to an anisotropy/multipole analysis~\cite{SiegalGaskins:2008ge,Lee:2008fm,Fornasa:2009qh}. This, in turn, may have important implications for DM cosmology; see the review
by T.~Bringmann in this issue.
\end{itemize}

\section{Conclusions}\label{conclusions}
The most challenging task for indirect dark matter searches is not to detect a few events, but to confidently identify those events as the products of dark matter annihilations. In particular, any signal must be separated from astrophysical backgrounds if it is to be reliably claimed to be a detection of dark matter. This is certainly true in the case of gamma-ray telescopes hoping to observe dark matter annihilations in the region of the Galactic Center (GC). The discovery of a bright TeV source of astrophysical origin at the GC by IACTs has changed the prospects for such searches considerably.  One strategy in light of this is to focus on different targets with lesser astrophysical background (such as dwarf galaxies). Another is to search for a signal in the ``noisy'' GC region, by taking advantage of the peculiar spectral and angular properties expected from dark matter annihilation products.

In this paper, we have reviewed the latter strategy, reporting on recent studies of the GC and inner Galaxy. Given the characteristics of this search, the LAT instrument on board of the Fermi satellite is better suited than existing IACTs. Fermi-LAT instrument will detect a number of astrophysical sources in the region of the sky around the GC, including the point sources previously identified by HESS and EGRET, and perhaps others. A diffuse gamma-ray background will also likely be present. Although predictions of Fermi's sensitivity are unavoidably limited by our incomplete knowledge of these backgrounds, we have shown that the spectral and angular differences between the signal and
backgrounds should be distinctive enough to allow one to separate signal from background over a significant region of the parameter space, at least for a sufficiently cusped dark matter profile (NFW-like or steeper).

In the optimistic case where dark matter annihilation products are identified by Fermi, then it may also be possible to measure or constrain the properties of the dark matter, including its mass, annihilation cross section, and spatial distribution.  It is unlikely that Fermi will determine the WIMP's mass with high precision, however. For example, for the case of a 100 GeV WIMP with an annihilation cross section of $3 \times 10^{-26}$ cm$^3$/s, distributed with an NFW halo profile, the mass could be determined to lie within approximately 50-300 GeV. In the same benchmark model, the inner slope of the dark matter halo profile could be determined to $\sim 10\%$ precision. The combination of several indirect detection channels will be crucial to both confirm such a detection, and to best constrain the WIMP's properties. 
On the other hand, it is not excluded that Fermi will lead to a radical revision of the present gamma-ray picture of the GC, revealing a more complicated zoo of astrophysical accelerators than envisaged in the present estimates. In the case where either the DM signal from the GC is too low or the  the background is too large/complex, a DM discovery in gamma rays is still possible by looking at the emission from an extended region in the inner halo with Fermi, or from other dark matter substructures with both Fermi-LAT and IACTs. In particular, the morphology and the spectral properties of the unresolved Galactic background at $E\lsim\,$GeV will be useful to optimize the angular and energy-cut templates for searches of the DM emission from an annulus of several tens of degrees around the GC.

\section*{Acknowledgments} 
PS would like to thank the Galileo Galilei Institute for Theoretical Physics for the hospitality and the INFN for partial support during the completion of this work. DH is supported in part by the Fermi Research Alliance, LLC under Contract No.~DE-AC02-07CH11359 with the US Department of Energy and by NASA grant NNX08AH34G. Finally, we thank Sergio Palomares Ruiz for noticing a typo in the v2 of the manuscript. 

\appendix
\section{The Case of Decaying Dark Matter}\label{DDM}

In the case of decaying dark matter, Eq.~(\ref{flux1}) is modified to
\begin{equation}
\Phi_\gamma=  \frac{\d N_{\gamma}}{\d E_{\gamma}} \frac{\Gamma}{4\pi\,m_X} \int_{\rm{los}} \rho(\ell,\Omega)\d \ell,
\label{fluxdec}
\end{equation}
where $\Gamma$ is the decay width (inverse lifetime) and the spectrum now refers to the photons generated in the decay process. Unlike with the cross section in the case of annihilating dark matter, one does not have any strong theoretical motivation for considering any particular lifetime for an unstable DM particle. In any case, arguments have been put forward justifying the typical range of the lifetimes needed for significant signatures in
astrophysics with $\sim$TeV mass particles and GUT-scale physics mediating the process (in analogy
with the expected proton decay in GUTs), see e.g.~\cite{Arvanitaki:2008hq}.
From the phenomenological point of view, there are a couple of points worth mentioning
regarding decaying DM candidates:
\begin{figure}[!htb]
\begin{center}
\epsfig{file=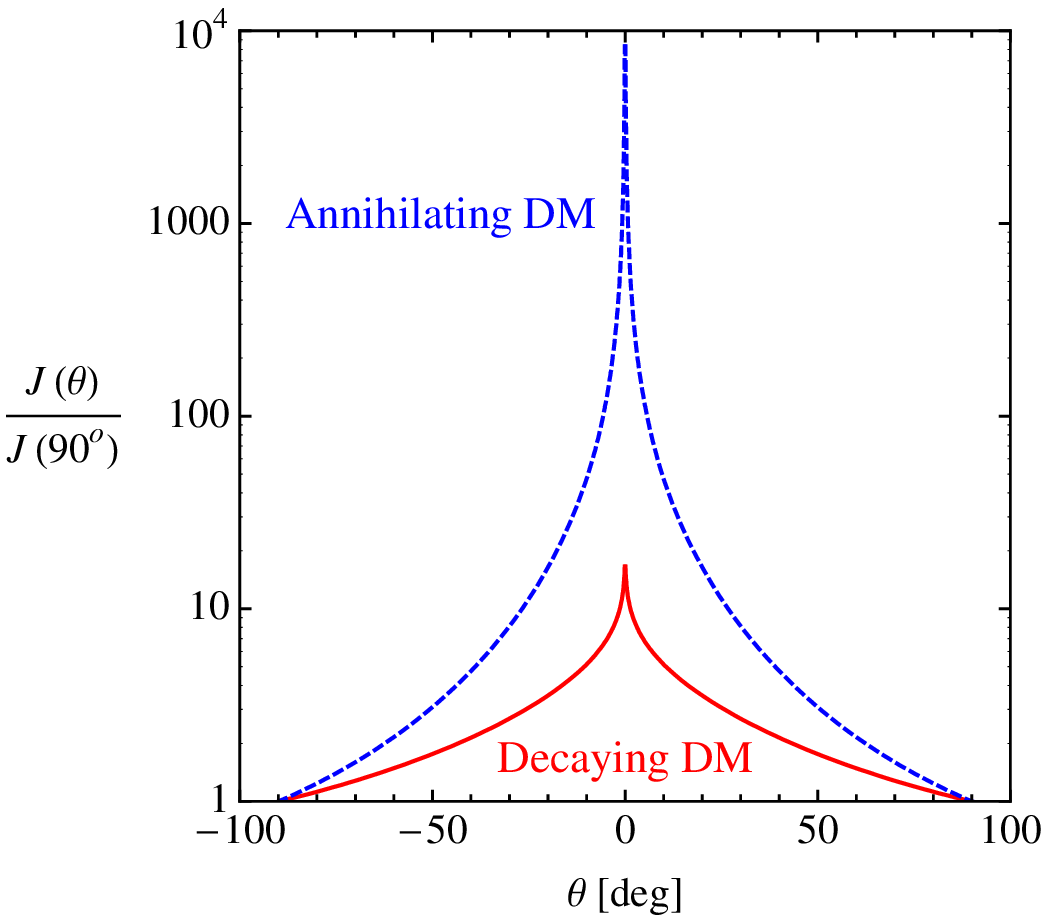,width=10cm}
\end{center}  
\caption{The angular profile of the gamma-ray signal as function of 
the angle, $\theta$, to the center of the galaxy for a NFW 
halo distribution for decaying DM (solid red line), 
compared to the case of self-annihilating DM (dashed blue line).
Both signals have been normalized to their values
at the galactic poles, $\theta = \pm 90^\circ$.
The central cusp is regularized by assuming in both cases an angular resolution of $0.1^\circ$.}
\label{fig:angular}
\end{figure}

\begin{itemize}
\item[I.] The DM distribution and the role of substructures in particular is of little importance in determining the level of the signal.
\item[II.] The angular distribution of the gamma-ray signal is very distinctive, and much flatter than the corresponding annihilation signal, as illustrated for a NFW profile in Fig.~(\ref{fig:angular}).
\end{itemize}

Should gamma rays be detected from DM, a comparison between the emission in the inner Galaxy and the emission at high latitude would immediately reveal the nature of the particle physics process (annihilation or decay) responsible for the emission~\cite{Bertone:2007aw}. Notice that this information is very difficult to extract with other cosmic ray probes.

\section{A Comment on the Factorization Assumption in Eq.~(\ref{flux1})}\label{entanglingEandAng}

Although the factorization between the particle physics term and astrophysical term in Eq.~(\ref{flux1}) is a useful approximation and valid in most practical cases, there are exceptions. More correctly, one should write
\begin{equation}
\Phi_{\gamma} (E_\gamma,\Omega)=\frac{\d N_{\gamma}}{\d E_{\gamma}} (E_\gamma)\frac{1}{8\pi m^2_X} \int_{\rm{los}}\left[\int\sigma (v_{r})\,v_{r}\,u({\bf v}_r){\rm d}^3{\bf v}_r\right] \rho^2(\ell,\Omega)\, \d \ell,
\label{flux3}
\end{equation}
where ${\bf v}_r$ is the relative velocity between the two particles (with $v_r\equiv |{\bf v}_r|$) and $u({\bf v}_r)$ its distribution
function (not necessarily isotropic), whose integral over ${\rm d}^3{\bf v}_r$ is normalized to unity.
The factorization assumed in Eq.~(\ref{flux1}) holds only if the integral in square brackets---which is nothing but $\sigv$---is independent of position. A sufficient condition for this is that $\sigma (v_{r})\,v_{r}$ is velocity-independent.  In general, the integral depends on the kinematical structure of the halo via the position-dependent velocity dispersion, anisotropy, etc. Then, both the astrophysical distribution of the DM and the particle physics contribute in determining the angular shape of the signal. 

One case in which the factorization is not valid can be found when the WIMP annihilations mainly through a $P$-wave process, such that $\sigma (v_{r})\,v_{r}\propto v_r^2$~\cite{Amin:2007ir}. This is a largely academic case, however, since whenever $P$-wave annihilation is dominant the
gamma-ray signal is expected to be suppressed. More interesting is the case in which non-perturbative processes lead to large ``Sommerfeld enhancements'' to the annihilation cross section at low velocities~\cite{Hisano:2004ds}.  This effect can be thought of as the distortion of the wave-function due to a relatively long-range
attraction between the WIMPs. In this case, a further
steepening of the signal towards the Galactic Center is expected, which in turn should ease the detection
of gamma rays from the inner Galaxy~\cite{Robertson:2009bh}.

\section*{References} 

\end{document}